\documentclass[12pt,letterpaper]{article}
\usepackage[margin=1.2in]{geometry}
\usepackage{setspace}

\usepackage{epsfig, amsthm, amsmath, amssymb, mathrsfs, latexsym, xpatch, xcolor, mathtools, verbatim, hyperref}
\usepackage[mathscr]{euscript}
\usepackage[titletoc]{appendix}
\usepackage[normalem]{ulem} 
\usepackage{tabularx}
\usepackage{stmaryrd} 
\usepackage[margin=1cm]{caption}
\usepackage{float} 
\usepackage[most]{tcolorbox}

\definecolor{cardinal}{rgb}{0.8, 0.0, 0.0}

\newcommand{\op}{\begin{itemize}}
\newcommand{\ed}{\end{itemize}}

\newcommand{\opp}{\begin{tcolorbox}[colback=white]}
\newcommand{\edd}{\end{tcolorbox}}

\newcommand{\ope}{\begin{enumerate}}
\newcommand{\ede}{\end{enumerate}}

\newcommand{\im}{\item}

\newcommand{\PP}{\mathbb{P}}

\title{Internalist Reliabilism in Statistics and Machine Learning: Thoughts on Jun Otsuka's {\em Thinking about Statistics}}
\author{Hanti Lin \\[0.5em] University of California, Davis \\ika@ucdavis.edu}

\date{To Appear in \\{\em The Asian Journal of Philosophy}}

\begin{document}

\maketitle

\begin{abstract} \noindent Otsuka (2023) argues for a correspondence between data science and traditional epistemology: Bayesian statistics is internalist; classical (frequentist) statistics is externalist, owing to its reliabilist nature; model selection is pragmatist; and machine learning is a version of virtue epistemology. Where he sees diversity, I see an opportunity for unity. In this article, I argue that classical statistics, model selection, and machine learning share a foundation that is reliabilist in an unconventional sense that aligns with internalism. Hence a unification under internalist reliabilism.
\end{abstract}


\section{Introduction}


Epistemologists of the more traditional variety have had little interaction with their counterparts in philosophy of science and even less with those in philosophy of data science, broadly construed to include statistics and machine learning. This is unfortunate, and perhaps even a scandal, because data science seems to be scientists' explicit attempt to develop an epistemology for their own inferential practices. Otsuka's book is the first monograph to take up the important task of addressing this gap, and thereby deserves attention from a broad audience in epistemology and philosophy of science.

Otsuka examines various distinctive approaches in data science and argues that they correspond to different camps in traditional epistemology. He develops a thought-provoking thesis: 
	\ope 
	\im[(i)] Bayesian statistics is internalist.
	\\[-2.2em]
	\im[(ii)] Classical (frequentist) statistics is externalist, owing to its reliabilist nature.
	\\[-2.2em]
	\im[(iii)] Frequentist model selection is pragmatist.
	\\[-2.2em]
	\im[(iv)] Machine learning aligns with virtue epistemology.
	\ede 
However, it seems to me that all these four approaches in data science are internalist, or at least can be interpreted as such. I even think that those mentioned in (ii)-(iv) fall under what could be termed {\em internalist reliabilism}, a combination that many traditional epistemologists might view as incoherent or even impossible.\footnote{But see Steup (2004) for a rare exception.} 

Before elaborating on my disagreement with Otsuka, let me briefly review the internalism-externalism debate in traditional epistemology.

\section{Internalism vs. Externalism}

Internalism was originally proposed as a thesis about the justification of belief rather than inference; it holds that all factors determining whether an agent's belief is justified ``reside within the agent'' (Otsuka 2023: 90). Internal factors include, for example, the beliefs one holds, the assumptions one is willing to take for granted, the propositions one treats as reasons for beliefs, and the deductive logical relations among propositions. In contrast, a paradigm example of an external factor is the actual reliability of a belief-producing process, which depends on whether the external world, so to speak, cooperates. The external world encompasses anything outside one's mental life, such as stars, apples, and even the neurons in one's brain.

While the above are relatively uncontroversial examples, the exact boundary between internal and external factors remains debated. The key point is that internalists focus exclusively on internal factors to ensure that one can in principle evaluate one's own beliefs and justify them {\em from within} one's first-person perspective. Externalists disagree, rejecting the need for such a requirement. For more on the nature of this debate, see the survey by Bonjour (2005).

The term `internalist' is most commonly used to describe theories of justified belief, but it can be applied more broadly. We can ask whether an inference method is justified for an agent, or whether an agent's adoption of a particular inference method is justified. If you believe that the answer depends solely on internal factors---those residing within the agent---you are an internalist about the justification of inference methods. Conversely, if you think that it depends on at least one external factor, you are an externalist.

In fact, most externalists seem to think that inference methods and other belief-producing procedures are the {\em primary} objects of assessment, with the assessment of beliefs being {\em derivative}. Consider this paradigm approach to externalism: 
	\opp 
	{\bf The Conventional Version of Reliabilism.} A person's belief is justified iff the method, procedure, or process by which that belief is produced is (in fact) highly reliable. 
	\edd 
The underlying idea can be understood as a two-step justification: high reliability justifies an inference method or belief-producing procedure, which, once justified, is able to justify the beliefs it produces. 

The first issue I want to address is whether classical statistics, as a theory for evaluating inference methods, is internalist or externalist. This is where I disagree with Otsuka. He argues that classical statistics is externalist, while I believe it is internalist---or at least can be naturally and plausibly interpreted as such.

I will first explain why it might appear to be tempting to view classical statistics as externalist before I argue that it is, in fact, internalist. I will start with examples from hypothesis testing and then proceed to other inference tasks, such as estimation and model selection.

\section{Hypothesis Testing}

Suppose that a scientist is testing a hypothesis $H_0$ with a prescribed sample size $n$. An inference method for this task, or a {\em test}, is a function that outputs a verdict---either ``Reject $H_0$'' or ``Don't''---whenever it receives a data sequence of the given length $n$. Classical statisticians widely accept a norm for assessing tests:\footnote
	{But perhaps this is only tacitly accepted, for statisticians rarely use the term `justified'.} 
	\opp 
	A test $T$ is justified (or one's adoption of test $T$ is justified) only if $T$ has a low significance level $\alpha$ (say 5\%).
	\edd 
But how are significance levels defined? This is the crux. Let me walk you through two alternative definitions.

\subsection{Informal Definition}

The following informal definition is quite common, being a primary source of the externalist impression of classical statistics. It is adopted by Otsuka (2023: 23), and can be found in many textbooks such as Rosner (2016: 213-214):
	\opp
	{\bf Informal Definition.} A test $T$ is said to have a low significance level (at level $\alpha$) iff the chance that $T$ erroneously rejects the tested hypothesis---known as the Type I error probability---is low (at most $\alpha$).
	\edd 
Notably, if the Type I error probability of a test exists, it is a physical chance---an objective property of some process in the actual world. Therefore, a test is automatically unjustified if it fails to have a low Type I error probability, irrespective of the first-person perspective of the scientist conducting the test. It does not matter whether the scientist can provide a good reason for believing that the Type I error probability of the test in use is high or low. Thus, classical hypothesis testing is rendered externalist. 


However, the above presentation is actually misleading. No such externalist interpretation is readily available once we turn to a more formal definition.

\subsection{Formal Definition}

The following definition is taken from a standard textbook by Casella \& Berger (2002). They define significance levels as follows (pp. 383, 385):
	\opp 
	{\bf Definition (Significance Level).} A test $T$ is said to achieve {\em significance level} $\alpha$ iff, for every $\theta$ in the given parameter space $\Theta$, if $\theta$ is in $H_0$ (that is, if $\theta$ is a possible world in which $H_0$ is true), then 
	$$
	\PP_{\theta} \big( \text{$M$ rejects $H_0$} \big) \;\le\;\alpha \,,
	$$
	where $\PP_\theta$ is the probability distribution determined by $\theta$ (i.e., the true probability distribution in world $\theta$). 
	\edd 
The parenthetical clauses are my interpretations, but they seem to be quite plausible. Like philosophers of language, mathematical statisticians formally represent a hypothesis or proposition as a set---the set of the possible ways in which it is true.

Now it should be clear what is required by a low significance level: the probability of a test's erroneous rejection of $H_0$ be kept uniformly low ($\le \alpha$) across all possible worlds in $\Theta$ where $H_0$ is true. This criterion is often conjoined with the following one (Casella \& Berger 2002: 388):  
	\opp
	{\bf Definition (Uniform Maximum Power).} A test is said to be {\em uniformly most powerful} at significance level $\alpha$ iff
		\ope 
		\im $T$ achieves significance level $\alpha$, 
		\im in each possible world $\theta \in \Theta$, if $\theta$ is not in $H_0$ (that is, if the tested hypothesis $H_0$ is false in possible world $\theta$), then $T$ is such that, for any alternative test $T'$ with the same significance level $\alpha$,
	\begin{eqnarray*}
	\PP_{\theta} \big( \text{$T$ rejects $H_0$} \big) 
	& \ge & 
	\PP_{\theta} \big( \text{$T'$ rejects $H_0$} \big) \,.
	\end{eqnarray*}
		\ede 
	\edd 
So, put intuitively, uniform maximum power requires that the probability of (correct) rejection attains maximum among all tests with a low significance level. This finishes a brief recap of the key formal definitions in the classical, Neyman-Pearson approach to hypothesis testing.

Here is the crux: these formal definitions refer to a set $\Theta$, called a parameter space. How should $\Theta$ be interpreted? 

\subsection{Interpretations of the Parameter Space $\Theta$}

In practice, the exact identity of $\Theta$ is highly content-dependent. Consider a scientist testing the hypothesis that the proportion of red marbles in an urn falls within a certain interval, under the background assumption that there are exactly 100 marbles in the urn, and under the usual assumption of IID (independent and identically distributed) data. In this context, the parameter space $\Theta$ is formally identified with the set of all (epistemically) possible proportions of red balls: 
	\begin{eqnarray*}
	\Theta &=& \left\{ \frac{a}{100}: a = 0, 1, 2, \dots, 100\right\} \,.
	\end{eqnarray*}
More precisely, $\theta$ represents the possible world in which the proportion of red marbles equals $\theta$ and the background assumptions (including IID data) are true, which determines a unique probability distribution $\PP_\theta$ (over possible data sequences)---the distribution true in world $\theta$. If the scientist's background assumptions are weaker---such as only assuming that there are no more than 100 marbles in the urn---the parameter space $\Theta$ must expand to accommodate more possible worlds. Thus, $\Theta$ can be reasonably identified with the set of possible worlds compatible with the background assumptions adopted in a context of inquiry. Under this interpretation, the Neyman-Pearson theory of hypothesis testing is internalist.

This internalist interpretation is applicable to frequentist statistics across the board, extending beyond hypothesis testing. Indeed:
	\opp 
	{\bf Frequentist Statistics as Reliabilism 1.} In frequentist statistics, inference methods are always assessed by criteria of a specific kind, namely {\em standards of reliability}, which examine the relevant \uline{reliability}$_\text{\,(i)}$ of an inference method in each of the possible worlds across a \uline{range $\Theta$}$_\text{\,(ii)}$. 
	\edd 
First focus on the underlined part (i): reliability. The relevant reliability in hypothesis testing is the probability of freedom from error, as we have seen above. In point estimation, the relevant reliability is the probability of producing a point estimate close to the true, unknown estimand. It is the use of standards of reliability that make frequentist statistics a reliabilist theory. 

But being reliabilist does {\em not} automatically imply being externalist. It depends on the underlined part (ii): the parameter space $\Theta$ as a domain of quantification. To be sure, frequentist statistics can be rendered externalist by restricting the parameter space $\Theta$ to a singleton containing only the actual world---whatever the actual world turns out to be, independent of anyone's background beliefs or assumptions. Then frequentist statistics becomes not just externalist, but also reliabilist in the conventional sense---it is all about reliability in the actual world. Yet such an externalist interpretation does not hold automatically: it relies on a substantive view of what $\Theta$ is.

In practice, the choice of $\Theta$ is highly context-dependent, as seen in the urn example, and $\Theta$ can be plausibly interpreted as the set of the possible worlds compatible with the background assumptions that one takes for granted in one's context of inquiry---in short, the set of the cases that {\em one deems possible}. This feature is crucial for serving a need of working scientists: namely, allowing them assess inference methods from within their first-person perspectives. I hasten to add that a first-person perspective can also be first-person {\em plural}: a group of scientists can work together to assess inference methods. The parameter space $\Theta$ can represent the common ground of those scientists---the assumptions commonly accepted within that group.

Therefore, under a plausible interpretation, frequentist statistics is reliabilist in an unconventional sense that aligns with internalism.

\section{Estimation}

Now, let me provide additional examples to illustrate the same point: frequentist statistics as internalist reliabilism. I will shift the focus from hypothesis testing to estimation. This transition will also help set the stage for our next topic: model selection and machine learning, which are closely connected to estimation.

Imagine a scientist aiming to estimate a certain quantity $\mu$. 
Suppose there is no prescribed sample size, possibly because this estimation project is ongoing, with the estimate updated as new data arrive. In this context, an estimator $\hat\mu$ is a function that produces a point estimate when given a data sequence of any finite length.

It would be great if an estimator $\hat\mu$ could come with a guarantee of a specific sample size $N$ that suffices for the desired reliability in estimation: a high probability of producing an estimate close to the unknown estimand. Here, a guarantee means a guarantee under the background assumptions that the scientist takes for granted, formalized by $\Theta$. This idea leads to the following criterion:
	\opp {\bf Definition (Uniform Consistency)}.
	An estimator $\hat\mu$ is called {\em uniformly consistent} iff, 
	\\[.8em]
	(i) for any level of high probability $1-\alpha$ less than one, 
	\\ (ii) for any level of tolerable error $\epsilon$ greater than zero, 
	\\ (iii) there exists a sample size $N$ such that,
	\\ (iv) for any possible world $\theta \in \Theta$,
	\\ (v) for any sample size $n \ge N$,
	\begin{eqnarray*}
	\PP_{\theta, n} \Big(  
	\text{$\hat\mu$ outputs an estimate $\epsilon$-close to the truth in world $\theta$} 
	\Big)
	& \ge & 1- \alpha \,,
	\end{eqnarray*}
	where $\PP_{\theta, n}$ denotes the probability distribution true in world $\theta$ over data sequences of length $n$. 
	\edd 
In some problem contexts, this standard is too high to be unachievable. For instance, consider the context in which one seeks to estimate the mean of a normal distribution with an {\em unknown} variance---hence a large parameter space $\Theta$. 
In this case, uniform consistency is unachievable, necessitating a shift to a lower standard, such as the following:
	\opp {\bf Definition (Pointwise Consistency)}. An estimator $\hat\mu$ is called {\em pointwise consistent} (or simply {\em consistent}) iff $\hat\mu$ satisfies the same condition as uniform consistency, except that the order of two quantifiers is swapped: (iii) `there exists $N$' and (iv) `for any $\theta \in \Theta$'.
	\edd 
Both uniform and pointwise consistency serve as standards of reliability, and an internalist interpretation is available owing to the quantification over $\Theta$---a point reiterated.

Yet there is an {\em additional} sense in which frequentist statistics is reliabilist. There is actually a hierarchy of standards of reliability, from high to low. A simple version looks like the following, where $\mathscr{D}$ is a desideratum added to pointwise consistency as a minimum qualification:
\opp 
$$\begin{array}{l}
	\text{Uniform Consistency}
\\
	\quad\quad\quad |
\\
	\text{Pointwise Consistency \,+\, $\mathscr{D}$}
\\
	\quad\quad\quad |
\\
	\text{Pointwise Consistency}
\\[1em]
	\text{(All are defined with respect to a space $\Theta$ of possible worlds)} 
\end{array}$$
\edd 
In fact, it seems to me that frequentist statistics operates under a somewhat tacit norm, which can be formulated as follows:\footnote
	{Lin (2022) formulates essentially the same norm in the context of formal learning theory.
	}
	\opp 
	{\bf Frequentist Statistics as Reliabilism 2 (An Achievabilist Norm).} An inference method is justified in a given context of inquiry only if it achieves the {\em highest} standard of reliability that is {\em achievable} (or proved to be achievable) for the inference problem undertaken in that context---pending the specification of the correct hierarchy of standards of reliability. 
	\edd 
This norm embodies a serious pursuit of reliability, {\em striving for the highest achievable standard of reliability}, which can be understood as an additional sense in which frequentists statistics is reliabilist. The above hierarchy is for point estimation; hypothesis testing has its own.\footnote
	{
	The following is the hierarchy discernible in the practice of hypothesis testing, where a low significance level is a minimum qualification:
$$\begin{array}{l}
	\text{Low Significance Level $\alpha$ \,+\, Uniform Maximum Power at Level $\alpha$}
\\
	\quad\quad\quad\quad\quad\quad |
\\
	\text{Low Significance Level $\alpha$ \,+\, $\mathscr{D}$}
\\
	\quad\quad\quad\quad\quad\quad |
\\
	\text{Low Significance Level $\alpha$}
\end{array}$$
	For an example of desideratum $\mathscr{D}$, see the discussion of unbiased tests in Casella \& Berger  (2002: pp. 387, 393).
	}

Returning to the hierarchy: an example of the additional desideratum $\mathscr{D}$ concerns the rate at which the outputs of an estimator converge to the truth---the {\em $\sqrt{n}$-rate of convergence}. This means, roughly, a rate of convergence as fast as the rate achieved in a classic case, in which the sample mean is used to estimate the mean of a normal distribution with an unknown variance. In practice, the hierarchy is much richer, incorporating a multiplicity of additional desiderata and their combinations.\footnote
	{For additional desiderata in point estimation, see Shao  (2003: chapters 3-5).}

The main point---frequentist statistics as internalist reliabilism---extends from point estimation to interval estimation almost immediately. When it is possible to achieve the high standard of uniform consistency in point estimation, we know how to achieve a correspondingly high standard in interval estimation: {\em prescribed confidence level with prescribed interval length} (Siegmund 1985: chapter 7). When it is possible to achieve the intermediate standard of pointwise consistency with the $\sqrt{n}$-rate of convergence in point estimation, we know how to achieve a weaker, asymptotic variant of the above standard in interval estimation, called {\em asymptotic confidence level} (Casella \& Berger  2002: section 10.4).

Thus, classical (frequentist) statistics---encompassing hypothesis testing, point estimation, and interval estimation---is both reliabilist and internalist across the board.


\section{Model Selection and Machine Learning}

The scope of internalist reliabilism extends further. As we will see shortly, the foundations of model selection and machine learning are actually extensions of the classical theory of estimation, which, as I have argued, is both internalist and reliabilist.

Imagine an economist aiming to predict the price of a house based on four factors: square footage, number of bedrooms, location, and age of the property. 
Or consider a computer scientist who wants to determine whether a given $1080 \times 1080$ image depicts a cat, based on the colors of its 1.2 million ($1080 \times 1080$) pixels. More generally, consider a scientist seeking to construct a function $f(x)$ to predict the value of $Y$ based on any given input $X = x$ (which could be a number or a vector). When $Y$ is binary, this task is known as {\em classification}; when $Y$ is real-valued, it is called {\em regression}. Classification is more frequently explored in machine learning, while regression has long been a traditional focus in statistics.

A very popular approach to classification and regression is {\em parametric modeling}. Examples include {\em polynomial models}, as shown below, where the $\beta_i$s are adjustable parameters:
	$$\begin{array}{ll}
	\text{Polynomial Model of Degree $1$:} & y =  \beta_0 + \beta_1 x
	\\
	\text{Polynomial Model of Degree $2$:} & y =  \beta_0 + \beta_1 x + \beta_2 x^2
	\end{array}$$
In general, a parametric model $\mathscr{M}$ is a class of functions that share a parametric form $f(x; \beta_1, \dots, \beta_k)$ with finitely many adjustable parameters $\beta_1, \dots, \beta_k$. When these parameter values are fixed, the model produces a specific function from $X$ to $Y$ (which can be visualized as a curve on the $XY$-plane). In machine learning, the most popular models are {\em neural network models}. Technical details aside, a neural network model is still a parametric model---a class of functions sharing a form $f(x; \beta_1, \dots, \beta_k)$, where the parameters $\beta_1, \dots, \beta_k$, known as {\em weights}, represent the strengths of signal transmission between neurons.

Given any dataset $D $ of $n$ points on the $XY$-plane as {\em training data}, a parametric model $f(x; \beta_1, \dots, \beta_k)$ is put to use as follows: the parameters $\beta_1, \dots, \beta_k$ are adjusted and fixed to produce a specific function that best fits the given dataset $D$. This resulting function is called a {\em fitted model}, also called a {\em predictor}, as it can be directly used for prediction.

Confronted with multiple models to choose from, we face an estimation problem: estimating the predictive power of each model. How is the predictive power defined? Suffices it to know that, under the usual assumption of IID data, a given model $\mathscr{M}$ has a well-defined {\em expected predictive accuracy} with respect to two factors: (i) a training sample size $n$, and (ii) a possible world $\theta$, which provides a probability distribution $\PP_\theta$ for defining expected values. So, when a model $\mathscr{M}$ and a set of $n$ data points on the $XY$-plane are provided as input, an estimator for the present purpose outputs an estimate of $\mathscr{M}$'s (actual) expected predictive accuracy for the training sample size $n$. 

The above raises some questions: What estimators may be used, and how should they be evaluated? These are foundational questions for model selection and machine learning. It is at this point that these two fields reconnect with the classical theory of estimation.

In the present context, it still makes sense to consider the classical standards for assessing point estimators, only with two differences. First, we need to be careful about the target of estimation: in the definitions of evaluative standards for point estimation, `the true value (to be estimated) in possible world $\theta$' needs to be replaced by `the expected predictive accuracy of the given model $\mathscr{M}$ with respect to possible world $\theta$ and training sample size $n$'. 

Here is the second difference: In the present context, the space $\Theta$ of possible worlds is intended to be very large, representing very weak background assumptions, such as the usual IID assumption and the assumption that the true probability distribution on the $XY$-plane is smooth. This makes $\{\PP_\theta: \theta\in\Theta\}$ a very large space of possible probability distributions, which do {\em not} share any particular parametric form. This is quite different from the classical theory in statistics, which traditionally employed strong background assumptions, often assuming that the true probability distribution takes a specific parametric form---hence the term `parameter space' associated with $\Theta$. Now that the background assumptions are weaker, I will refer to $\Theta$ directly as a space of possible worlds rather than a parameter space to avoid confusion with the parameters $\beta_1, \dots, \beta_k$ in a model.

Let's return to the task of estimating the expected predictive accuracies of models. This was pioneered by Akaike (1973), who proved that, under certain background assumptions, the AIC estimator meets at least a standard called {\em asymptotic unbiasedness}.\footnote 
	{	
	Asymptotic unbiasedness means that, as the sample size increases indefinitely, the expected value of the estimate converges to the true estimand in each possible world in $\Theta$. If, furthermore, the variance converges to zero in each possible world in $\Theta$, then pointwise consistency is achieved.
	} 
Although asymptotic unbiasedness is a relatively low standard---lower and weaker than pointwise consistency, as shown in the following hierarchy---Akaike's theorem marks the beginning of an important research program.
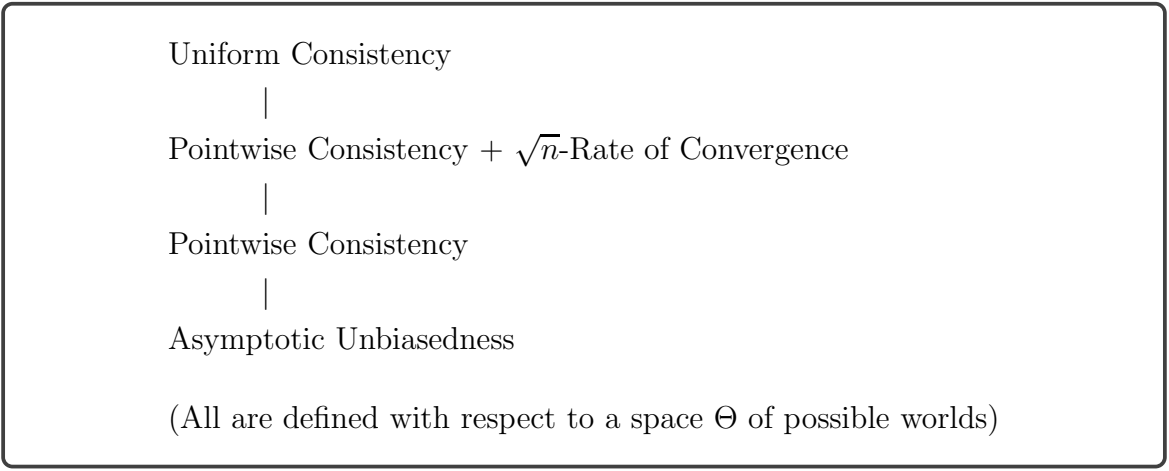
\opp 
$$\begin{array}{l}
	\text{Uniform Consistency}
\\
	\quad\quad\quad |
\\
	\text{Pointwise Consistency + $\sqrt{n}$-Rate of Convergence}
\\
	\quad\quad\quad |
\\
	\text{Pointwise Consistency}
\\
	\quad\quad\quad |
\\
	\text{Asymptotic Unbiasedness}
\\[1em]
	\text{(All are defined with respect to a space $\Theta$ of possible worlds)} 
\end{array}$$
\edd 
The question left by Akaike is whether it is possible to achieve a higher standard. Progress has been made in climbing the hierarchy, thanks to a series of breakthroughs since 2020 (Austern et al. 2020, Bayle et al. 2020, Wager 2020, Li 2023, and Bates et al. 2024). We now know that it is possible to design a single estimator that achieves at least the second standard---pointwise consistency with the $\sqrt{n}$-rate of convergence---for a broad range of models, extending well beyond polynomial models and neural network models.\footnote{
	For a list of the models covered, see Li (2023: pp. 497-499). A note on Li's terminology: what we call models here are called estimators by Li. Indeed, a model can be understood to use training data to estimate an unknown curve on the $XY$-plane.}
Moreover, this has been shown to be achieved by certain versions of cross validation, specifically $K$-fold cross validation, where $K$ is any positive integer held constant. 

To sum up: Although the fields of model selection and machine learning are relatively new, their theoretical foundation aligns with the classical theory of estimation. This foundation addresses specific needs: the need to estimate the expected predictive accuracies of the models under consideration and, consequently, the (higher-order) need to evaluate estimators, such as the AIC estimator and the cross-validation estimator. These estimators are assessed in the same style as in classical estimation: an internalist evaluation sensitive to background assumptions and framed in terms of standards of reliability. It employs the same hierarchy of standards and is guided by the same pursuit of the highest achievable standard. Thus, the underlying philosophy remains internalist reliabilism. The main difference is that when the focus shifts from the classical theory to model selection and machine learning, the target of estimation becomes less familiar, and the mathematics more complex. Yet, it remains internalist reliabilism throughout.

\section{Closing}

While Otsuka maintains that classical statistics is a form of externalist epistemology because it is reliabilist, I hold a different view. I have argued that classical statistics is most plausibly interpreted as internalist reliabilism.

Otsuka also contends that model selection is a form of pragmatist epistemology, and that machine learning aligns with virtue epistemology. I agree, but this diversity should not be overemphasized. As explained above, model selection and machine learning actually have a common foundation shared with classical statistics: a frequentist foundation, which can be interpreted as both internalist and reliabilist. Where Otsuka sees diversity, I see an opportunity for unity---a unification under internalist reliabilism. 

My disagreement with Otsuka on those finer points should not distract you from the bigger picture: Otsuka's admirable efforts to promote a great tradition in epistemology. It is the tradition of addressing epistemological questions by perusing the most exciting sciences of one's time. For Plato, the most exciting science was Euclidean geometry: he reflected on how knowledge is gained in Euclidean geometry (as in {\em Meno}), and used it as a model for acquisition of ethical knowledge (as in {\em The Republic}). In Kant's time, the leading science was Newtonian physics, to which he contributed early in his career, and his {\em Critique of Pure Reason} sought to establish an epistemological foundation of Newtonian physics along with the (Euclidean) geometry it presupposes. Today, the most exciting science is data science, broadly construed to conclude statistics and machine learning. Otsuka's book is the first monograph that systematically and comprehensively examine data science in relation to some of the deepest issues in epistemology. This article represents my attempt to follow his lead.



\section*{Acknowledgements}

I am indebted to Jun Otsuka, I-Sen Chen, and Konstantin Genin for their valuable discussions.

\section{References}

\begin{description}
\im Akaike, H. (1973). Information theory and an extension of the Maximum Likelihood Principle. In B. N. Petrov, \& F. Csaki (Eds.), \textit{Proceedings of the 2nd International Symposium on Information Theory} (pp. 267-281). Budapest: Akademiai Kiado.

\im Austern, M. \& Zhou, W. (2020). Asymptotics of cross-validation. arXiv preprint
arXiv:2001.11111.

\im Bates, S., Hastie, T., \& Tibshirani, R. (2024). Cross-Validation: What does it estimate and how well does it do it?. \textit{Journal of the American Statistical Association}, \textit{119}(546), 1434-1445.

\im Bayle, P., Bayle, A., Janson, L., \& Mackey, L. (2020). Cross-validation confidence intervals for test error. \textit{Advances in Neural Information Processing Systems}, \textit{33}, 16339-16350.

\im Bonjour, L. (2005). Internalism and externalism. In P.K. Moser (Ed.), \textit{The Oxford Handbook of Epistemology} (pp. 234-263). Oxford University Press.

\im Casella, G \& Burger, R. (2002). \textit{Statistical inference}, 2nd Edition. Duxbury Press.

\item Li, J. (2023). Asymptotics of K-fold cross-validation. {\em Journal of Artificial Intelligence Research}, {\em 78}, 491-526.

\im Lin, H. (2022). Modes of convergence to the truth: Steps toward a better epistemology of induction. \textit{The Review of Symbolic Logic}, \textit{15}(2), 277-310.


\im Otsuka, J. (2023). \textit{Thinking about statistics: The philosophical foundations}. Routledge.
 
\im Rosner, B. A. (2006). \textit{Fundamentals of biostatistics}. Belmont, CA: Thomson-Brooks/Cole.

\im Siegmund, D. (1985). \textit{Sequential analysis: tests and confidence intervals}. Springer.

\im Shao, J. (2003). \textit{Mathematical statistics}.  Springer.

\im Steup, M. (2004). Internalist Reliabilism. \textit{Philosophical Issues}, \textit{14}, 403-425.

\im Wager, S. (2020). Cross-Validation, risk Estimation, and model Selection: Comment on a Paper by Rosset and Tibshirani. \textit{Journal of the American Statistical Association}, \textit{115}(529), 157-160.
\end{description}

\end{document}